\documentclass[aps,prd,preprint,nofootinbib]{revtex4}
\usepackage{amsfonts}
\usepackage{mathrsfs}
\usepackage{graphicx}
\usepackage{amsmath}
\usepackage{amssymb}
\usepackage{multirow}
\usepackage{subfigure}
\usepackage{epsfig}
\usepackage{graphicx}
\usepackage{booktabs}
\usepackage{array}
\usepackage{tabularx}
\usepackage{slashed}
\usepackage{bbm}

\allowdisplaybreaks[4]

\graphicspath{{fig/}{dia/}} \DeclareGraphicsExtensions{.eps}
\begin{document}
\baselineskip 20pt
\title{The trace amplitude method and its application to the NLO QCD calculation}

\author{\vspace{1cm} Zi-Qiang Chen$^{1}$\footnote[2]{
chenziqiang13@mails.ucas.ac.cn}  and Cong-Feng
Qiao$^{1,2}$\footnote[1]{qiaocf@ucas.ac.cn, corresponding author} \\}

\affiliation{$^1$ School of Physics, University of Chinese Academy of
Sciences, Yuquan Road 19A, Beijing 100049\\
$^2$ CAS Key Laboratory of Vacuum Physics, Beijing 100049, China\vspace{0.6cm}}

\begin{abstract}

The trace amplitude method (TAM) provides us a straightforward way to calculate the helicity amplitudes with massive fermions analytically.
In this work, we review the basic idea of this method, and then discuss how it can be applied to next-to-leading order (NLO) quantum chromodynamics (QCD) calculations, which has not been explored before.
By analyzing the singularity structures of both virtual and real corrections, we show that the TAM can be generalized to NLO QCD calculations straightforwardly, the only caution is that the unitarity should be guaranteed.
We also present a simple example to demonstrate the application of this method.

\vspace {5mm} \noindent {PACS number(s): 12.38.--t, 12.38.Bx}
\end{abstract}

\maketitle

\section{Introduction}
In high energy colliders like the Large Hadron Collider (LHC), processes with multi-particle final states are of great important in signal and background analysis.
To describe these processes at the same precision level as the experimental measurements, one has to calculate the cross sections to at least next-to-leading order (NLO).
However, the NLO calculations for multi-particle processes are very challenging: as the number of external particles increases, both the number of Feynman diagrams and the computational difficulty of each diagram grows rapidly.
The conventional amplitude squaring approach (CAS), i.e. squaring the Feynman amplitude, summing over the spins of external states, and taking the trace of each possible fermion string loop, is proved to be tedious and time consuming when the number of external particles is more than 5.
Another drawback of this approach is that it loses the spin information of final state particles, which is attainable in present experimental measurements.

An alternative approach is to compute the helicity amplitude explicitly, then the amplitude squaring and polarization summation can be performed easily during the numerical evaluation. The development of this approach has experienced a long history.
Many techniques have been developed for calculating the tree- \cite{DeCausmaecker:1981jtq,Berends:1981uq,Nam:1983gt,Kleiss:1985yh, Xu:1986xb,Parke:1986gb,Berends:1987me,Chang:1992bb,Yehudai:1992rt, Ballestrero:1994jn,Vega:1995cc,Bondarev:1997kf,Andreev:2001se,Qiao:2003ue,Cachazo:2004kj,Britto:2004ap,Schwinn:2005pi} and loop- \cite{Bern:1991aq,Bern:1994zx,Bern:1994cg,Brandhuber:2004yw,Luo:2004ss, Bena:2004xu,Quigley:2004pw,Bedford:2004py,Britto:2004nc,Roiban:2004ix,Bern:2005hs,Bidder:2005ri} level helicity amplitudes.
It should be noted that literatures on this subject are vast, and to give a complete survey of them is beyond the scope of this paper.
For reviews, see for instance Refs. \cite{Mangano:1990by,Dixon:1996wi,Bern:2008ef,Elvang:2013cua,Dixon:2013uaa}.

For processes of fermion production or decays, Feynman amplitudes incorporate one or more open fermion line, which can be expressed as
\begin{equation}
\bar{U}(p_1,\lambda_1)\Gamma U(p_2,\lambda_2)={\rm tr}[\Gamma U(p_2,\lambda_2)\otimes\bar{U}(p_1,\lambda_1)],
\label{eq_fline}
\end{equation}
where $p_1$ and $p_2$ are the momenta of the external fermions, $\lambda_1$ and $\lambda_2$ denote their polarization states; $\Gamma$ stands for the string of Dirac gamma matrices between the spinors; $U(p,\lambda)$ stands for either fermion spinor $u(p,\lambda)$ or anti-fermion spinor $v(p,\lambda)$.
In $4$-dimensional spinor space, the spinor product $U(p_2,\lambda_2)\otimes\bar{U}(p_1,\lambda_1)$ can be re-expressed by basic Dirac gamma matrices through different ways \cite{Nam:1983gt,Kleiss:1985yh,Chang:1992bb,Yehudai:1992rt,Ballestrero:1994jn,Vega:1995cc,Bondarev:1997kf,Andreev:2001se,Qiao:2003ue}, then the trace in Eq. (\ref{eq_fline}) can be evaluated straightforwardly.
For convenience, we call this method the trace amplitude method (TAM) hereafter.
The TAM is different from the so called helicity amplitude method (HAM), which has been proposed in Refs. \cite{Kleiss:1985yh,Xu:1986xb} and generalized to NLO in Refs. \cite{Bern:1991aq,Bern:1994zx}.
These two methods are complementary to each other: the results obtained from the HAM is more compact, while the TAM is more transparent to beginners and more convenient for a realization on a computer algebra system.
Although the TAM has been proposed for a long time, its validity in higher-order calculations has not been discussed before.
Considering the fact that the NLO corrections are usually important in phenomenological study, in this work we discuss the generalization of  TAM to NLO QCD calculation.

The rest of the paper is organized as follows.
In Sec. II, we review some basic formulas used to derive the TAM;
In Sec. III, we analyze different types of singularities encountered in NLO QCD calculations;
In Sec. III, we present a scheme that enable the application of TAM in NLO QCD calculation;
In Sec. IV, illustrative examples, the NLO QCD corrections to $g+g\to t+\bar{t}$ and $q+\bar{q}\to t+\bar{t}$ processes, are presented.
The last section is reserved for a summary.

\section{Spinor product}
The key ingredient of the TAM is to re-express the spinor product $U(p_2,\lambda_2)\otimes\bar{U}(p_1,\lambda_1)$ by basic Dirac gamma matrices.
This re-expression can be done through various means, such as constructing the transformation matrix between spinors with different momenta and polarization states \cite{Nam:1983gt}, introducing auxiliary vectors \cite{Kleiss:1985yh,Chang:1992bb,Ballestrero:1994jn,Andreev:2001se,Qiao:2003ue}, making the use of orthogonal basis of the $4$-dimensional spinor space \cite{Yehudai:1992rt}, making the use of the Bouchiat-Michel identity \cite{Vega:1995cc,B-MIdentity}, etc.
In fact, as revealed in Ref. \cite{Bondarev:1997kf}, in fact all these approaches may attribute to the same mathematical scheme.
In this section, we follow the lines of auxiliary vector approach, and present some basic formulas of TAM.

Consider a (anti)fermion with momentum $p$, polarization vector $s$, and mass $m$, the on-shell and polarization conditions require that
\begin{equation}
p^2=m^2,\quad s^2=-1,\quad p\cdot s=0.
\end{equation}
The corresponding spinor can be defined as the common eigenstate of the two commuting operators $\slashed{p}$ and $\gamma_5\slashed{s}$:
\begin{align}
&\slashed{p}U_s(p,\lambda)=M U_s(p,\lambda),\label{eq_diracEq1} \\ 
& \gamma_5\slashed{s}U_s(p,\lambda)=\lambda U_s(p,\lambda).
\label{eq_diracEq2}
\end{align}
Here, for fermion $U_s(p,\lambda)=u_s(p,\lambda)$, $M=m$; for antifermion $U_s(p,\lambda)=v_s(p,\lambda)$, $M=-m$; $\lambda=\pm 1$ denote the two different polarization states.

The massive spinor $U_s(p,\lambda)$ can be constructed with massless spinor. By introducing two auxiliary vectors that fulfil the conditions
\begin{equation}
k_0^2=0,\quad k_1^2=-1,\quad k_0\cdot k_1=0,
\end{equation}
one can construct a massless spinor $w(k_0,\lambda)$ in light of the light-like vector $k_0$, satisfying
\begin{align}
&\slashed{k}_0w(k_0,\lambda)=0,\\
&\gamma_5 w(k_0,\lambda)=\lambda w(k_0,\lambda).
\end{align}
From above two equations we have 
\begin{align}
&w(k_0,\lambda)\bar{w}(k_0,\lambda)=\frac{1+\lambda \gamma_5}{2}\slashed{k}_0,\nonumber \\
&\slashed{k}_1w(k_0,\lambda)=\lambda w(k_0,-\lambda).
\label{eq_masslessPro}
\end{align}
Here the relative phase between $w(k_0,+)$ and $w(k_0,-)$ may be fixed by $k_1$.

With the massless spinor $w(k_0,\lambda)$, the massive spinor $U_s(p,\lambda)$ can be expressed as
\begin{equation}
U_s(p,\lambda)=\frac{(\slashed{p}+M)(1+\slashed{s})}{2\sqrt{k_0\cdot (p+M s)}}w(k_0,-\lambda) ,
\label{eq_construct}
\end{equation}
which satisfies \eqref{eq_diracEq1} and \eqref{eq_diracEq2}.
The normalization factor here is fixed by the fermion spin sum relation
\begin{equation}
 \sum_{\lambda}U_s(p,\lambda)\bar{U}_s(p,\lambda)=\slashed{p}+M.
 \label{eq_fspinsum}
 \end{equation}
Combine Eqs. \eqref{eq_masslessPro} and \eqref{eq_construct}, one can readily get the desired spinor product:
\begin{equation}
U_{s_1}(p_1,\lambda_1)\otimes\bar{U}_{s_2}(p_2,\lambda_2)=\frac{({\slashed p}_1+M_1)(1+{\slashed s}_1)\Lambda(\lambda_1,\lambda_2){\slashed k}_0(1+{\slashed s}_2)({\slashed p}_2+M_2)}{8\sqrt{k_0\cdot(p_1+M_1s_1)}\sqrt{k_0\cdot(p_2+M_2s_2)}},
\label{eq_sPro}
\end{equation}
with
\begin{align}
&\Lambda(\lambda,\lambda)=1-\lambda\gamma_5,\nonumber \\
&\Lambda(\lambda,-\lambda)=\slashed{k}_1(\lambda+\gamma_5).
\label{eq_lamga5}
\end{align}

The polarization vector of a fermion can be expressed through the momentum of the fermion as
\begin{equation}
s=\frac{(p\cdot q) p-m^2 q}{m\sqrt{(p\cdot q)^2-m^2q^2}},
\end{equation}
where $q$ can be an arbitrary vector except for those paralleling to momentum $p$. 
For $q=(1,\vec{0})$, the polarization vector is found to be $s=(\frac{|\vec{p}|}{m},\frac{E}{m}\frac{\vec{p}}{|\vec{p}|})$, which indicates that the corresponding spinor is in helicity eigenstate.
In computation, it is more convenient to take $q=\frac{M}{m} k_0$, the so called  Kleiss-Stirling (KS) \cite{Kleiss:1985yh} polarization basis.
In this basis, Eq. \eqref{eq_sPro} can be simplified to
\begin{equation}
U_{\rm KS}(p_1,\lambda_1)\otimes\bar{U}_{\rm KS}(p_2,\lambda_2)=\frac{({\slashed p}_1+M_1)\Lambda(\lambda_1,\lambda_2){\slashed k}_0({\slashed p}_2+M_2)}{4\sqrt{k_0\cdot p_1}\sqrt{k_0\cdot p_2}}.
\label{eq_sProS}
\end{equation}
Note, in phenomenological study, other choices of polarization basis may lead to certain convenience.
The transformation rule between spinors in different polarization basis can be obtained by taking an explicit representation for Dirac matrices.

In general, arbitrary vectors $k_0$ and $k_1$ will cause the result of the amplitude extra complication.
To avoid this, in actual computation, one may either specify $k_0$ and $k_1$ explicitly, or construct $k_0$ and $k_1$ with external momenta, as demonstrated in Ref. \cite{Chang:1992bb}.

\section{Singularity structure of NLO QCD calculation}

In this section, we analyze the singularity structure of NLO QCD calculation.
The dimensional regularization with space-time dimension $D=4-2\epsilon$ is used to regularize both ultraviolet (UV) and infrared (IR) singularities.
Although the results are well known \cite{Kunszt:1994np,Catani:2000ef}, we discuss them in detail for a twofold reason: (i) they are essential to the NLO generalization of TAM, and (ii) we provide a new perspective on this subject.
Specifically, in Ref. \cite{Catani:2000ef}, the singular terms of the virtual loop corrections are derived from that of the real corrections, by exploiting the fact that the IR singularities of the virtual and real corrections cancel each other. While here, we derive these singular terms through direct loop integral analysis, and show that they are exactly canceled by their counterparts in real corrections.

\subsection{Singular terms in virtual corrections}
\label{sec_sinV}
The one-loop virtual corrections contain UV and IR singularities, which appear as $\frac{1}{\epsilon^n}-$pole under the dimensional regularization.
In a renormalizable theory like QCD, UV singularities are contained in the diagrams or subdiagrams with a small number of external legs, and can be removed by renormalization procedure.
In renormalized perturbation theory, the renormalized UV-finite one-loop amplitude $\tilde{\mathcal{M}}^{\rm loop}$ is defined as
\begin{equation}
\tilde{\mathcal{M}}^{\rm loop}=\mathcal{M}^{\rm loop}+\mathcal{M}^{\rm CT},
\label{eq_UVterm}
\end{equation}
where $\mathcal{M}^{\rm CT}$ denotes the amplitudes of counterterms.

To study the IR singularity structure of $\tilde{\mathcal{M}}^{\rm loop}$, we use lightcone gauge, where the gluon propagator is
\begin{equation}
D_{\mu\nu}^{ab}(p)=\delta^{ab}\frac{i\Pi_{\mu\nu}(p)}{p^2+i\varepsilon},
\end{equation}
with
\begin{equation}
\Pi_{\mu\nu}(p)=-g_{\mu\nu}+\frac{r_\mu p_\nu+r_\nu p_\mu}{r\cdot p},
\end{equation}
where $r$ is a light-like vector.
Lightcone gauge is a physical gauge, means a sum over physical transverse polarization states while the gluon is on its mass shell:
\begin{equation}
\Pi_{\mu\nu}(p)\overset{p^2=0}{\longrightarrow}\sum_{i=1,2}\epsilon_\mu^{(i)}(p,r)\epsilon_\nu^{(i)*}(p,r).
\end{equation}

For one-loop amplitude without soft or mutually collinear external lines, soft singularities originate from the exchange of soft gluon between two on-shell legs.
To isolate the soft singularities, we impose a cutoff $\delta_0$ to all components of loop momentum $k$, that is
\begin{equation}
|k^\mu|<\delta_0\ll \text{particle masses or other kinematic scales}.
\label{eq_softregion}
\end{equation}
This region will be referred as soft region.

The structure of external leg attached by a soft gluon can be approximated as
\begin{equation}
\vcenter{\hbox{\includegraphics[scale=0.5]{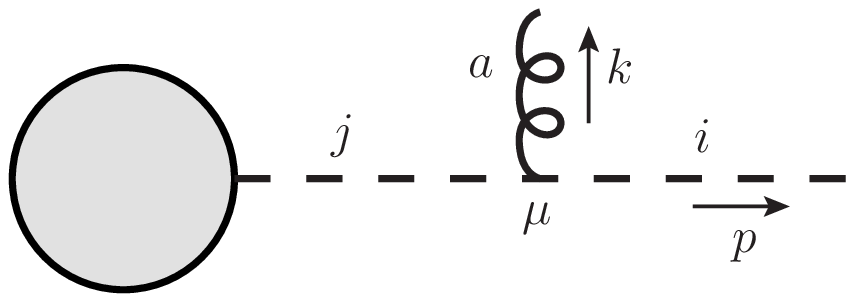}}}\simeq-g_s T^a_{c_i c_j}\frac{1}{k^2+2p\cdot k}\left(2p^\mu \vcenter{\hbox{\includegraphics[scale=0.5]{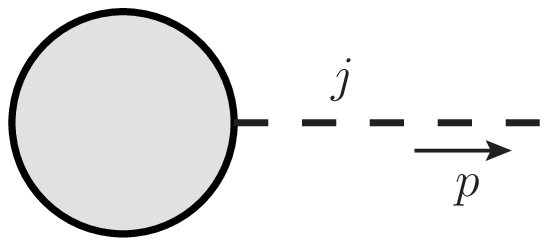}}} + \mathcal{O}(|k|)\right),
\label{eq_eikonal}
\end{equation}
with
\begin{equation}
T^a_{c_i c_j}=
\begin{cases}
t^a_{c_i c_j}, i=\text{outgoing quark or incoming antiquark}\\
-t^a_{c_j c_i},i=\text{outgoing antiquark or incoming quark}\\
-if^{a c_i c_j},i=\text{gluon}
\end{cases} .
\end{equation}
Here the dashed line denotes either quark (massive or massless) or gluon, $c_i$ denotes the color index of parton $i$, $t^a_{c_ic_j}$ and $f^{ac_ic_j}$ are the generators of $SU(3)$ fundamental and adjoint representation respectively.
By default, the momentum of parton $i$ is defined as outgoing.
For incoming case, one should take the replacement $p\to -p$.

For the structure of two external legs connected by one soft gluon, we have
\begin{align}
\vcenter{\hbox{\includegraphics[scale=0.5]{softloop.eps}}}\simeq&ig_s^2 T^a_{c_ic_{i^\prime}} T^a_{c_jc_{j^\prime}}\mu^{4-D}\int\limits_{|k^\mu|<\delta_0} \frac{d^Dk}{(2\pi)^D}\frac{1}{k^2(k^2+2k\cdot p_i)(k^2-2k\cdot p_j)}\\ \nonumber
&\times\left(4p_i\cdot \Pi(k)\cdot p_j\vcenter{\hbox{\includegraphics[scale=0.5]{softloopB.eps}}}+\mathcal{O}(|k|)\right) .
\label{eq_softijeikonal}
\end{align}
Here, all the three propagators have poles in the region $|k^\mu|<\delta_0$.
However, at NLO, only the poles of $1/k^2$ are concerned.
For the case where both partons are incoming or outgoing, the poles of $1/(k^2+2k\cdot p_i)$ and $1/(k^2-2k\cdot p_j)$ lead to pure imaginary singular terms\footnote{These terms can be obtained through the Cutkosky rules \cite{Cutkosky:1960sp}.}, which eventually canceled each other between $\tilde{\mathcal{M}}^{\rm loop}(\mathcal{M}^{\rm tree})^*$ and $(\tilde{\mathcal{M}}^{\rm loop})^*\mathcal{M}^{\rm tree}$.
For the case where one parton is outgoing while another is incoming, no imaginary singularities arise.
In fact, the poles of $1/(k^2+2k\cdot p_i)$ and $1/(k^2-2k\cdot p_j)$ are located in the region where $k^+k^-\ll |\vec{k}_T|^2$  \footnote{Here we work in the $p_i+p_j$ center-of-mass frame.
We take the lightcone coordinates so that the momenta are $p_i=(p_i^+,m_i^2/(2p_i^+),\vec{0}_T)$, $p_j=(m_j^2/(2p_j^-),p_j^-,\vec{0}_T)$, $k=(k^+,k^-,\vec{k}_T)$.}, and we can perform contour deformations on both $k^+$ and $k^-$ to get out of this region \cite{Collins:2011zzd}.
Note, since we work in lightcone guage, the singularities at $r\cdot k=0$ may obstruct the contour deformations.
In our simple case, this can be overcome by appropriate choice of $r$.
For example, we may choose a generic $r$ that do not parallel to any $p_i$.

When both partons $i$ and $j$ are massive, we can deform the integral to a contour where all components of $k$ are comparable.
Then the asymptotic behavior of loop momentum is $|k^\mu|\sim \kappa^2$ as $\kappa\to 0$.
Thus we can neglect the $k^2$ term compared to $k\cdot p_i$ or $k\cdot p_j$ (eikonal approximation).
It can also be seen that the $\mathcal{O}(|k|)$ term does not contribute to soft singularity, as it leads to terms scaling like $\kappa^2$ or higher.
When either or both of $i$ and $j$ are massless, there is an overlapping soft-collinear region, where the scaling behavior of $k$ is $|k^+|\sim \kappa$, $|k^-|\sim \kappa^3$ and $|\vec{k}_T|\sim \kappa^2$ (or $|k^+|\sim \kappa^3$, $|k^-|\sim \kappa$ and $|\vec{k}_T|\sim \kappa^2$).
It can be seen that the eikonal approximation still hold in this region.
Thus for both massive and massless cases, we have
\begin{align}
\vcenter{\hbox{\includegraphics[scale=0.5]{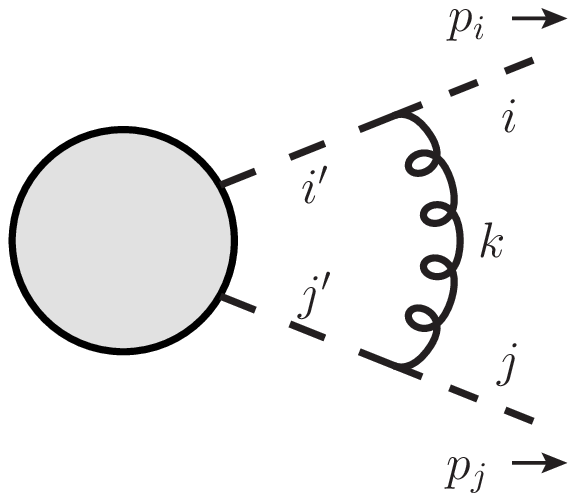}}}\overset{\rm soft}{\sim}-ig_s^2 T^a_{c_ic_{i^\prime}} T^a_{c_jc_{j^\prime}} \mu^{4-D}\int\limits_{|k^\mu|<\delta_0} \frac{d^Dk}{(2\pi)^D}\frac{p_i\cdot \Pi(k)\cdot p_j}{k^2(k\cdot p_i)(k\cdot p_j)} \vcenter{\hbox{\includegraphics[scale=0.5]{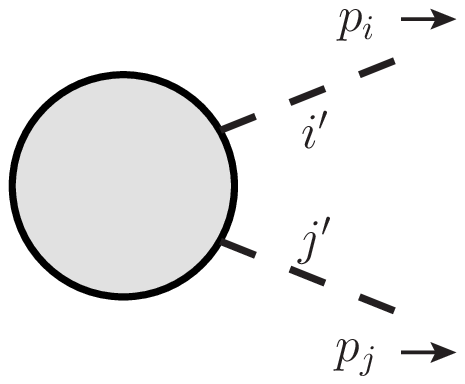}}}&\nonumber \\
\overset{\rm soft}{\sim}-g_s^2 T^a_{c_ic_{i^\prime}}T^a_{c_jc_{j^\prime}} \mu^{4-D}\int\limits_{|\vec{k}|<\delta_0} \frac{d^{D-1}k}{2k_0(2\pi)^{D-1}}\frac{p_i\cdot \Pi(k)\cdot p_j}{(k\cdot p_i)(k\cdot p_j)}\bigg|_{k_0=|\vec{k}|} \vcenter{\hbox{\includegraphics[scale=0.5]{softLoopB.eps}}}&\ .
\end{align}
Here, the symbol ``$\overset{\rm soft}{\sim}$'' denotes that the real part of soft singular terms on each side are equal.

Besides soft gluon exchange between external legs, soft singularities also come from on-shell renormalization constants.
The corresponding terms can be re-expressed as self-energy insertions to external lines.
For this case, we have (see Appendix A for detailed derivation):
\begin{align}
\frac{1}{2}\vcenter{\hbox{\includegraphics[scale=0.5]{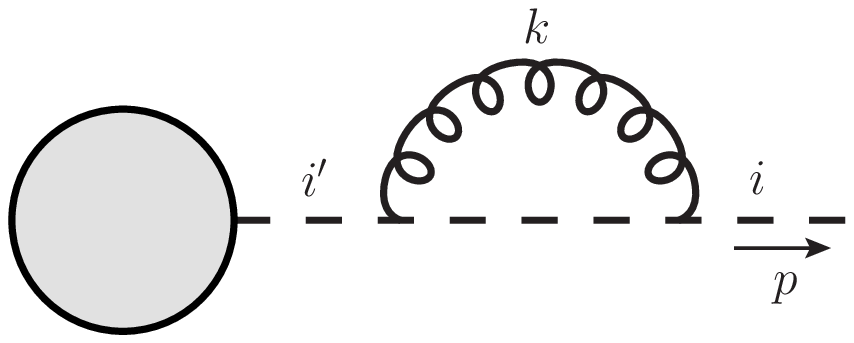}}}\overset{\rm soft}{\sim}&-\frac{1}{2}g_s^2 T^a_{c_ic_{i^{\prime\prime}}}T^a_{c_{i^{\prime\prime}}c_{i^\prime}} \mu^{4-D}\int\limits_{|\vec{k}|<\delta_0} \frac{d^{D-1}k}{2k_0(2\pi)^{D-1}}\frac{p_i\cdot \Pi(k)\cdot p_i}{(k\cdot p_i)^2}\bigg|_{k_0=|\vec{k}|} \nonumber \\
&\times \vcenter{\hbox{\includegraphics[scale=0.5]{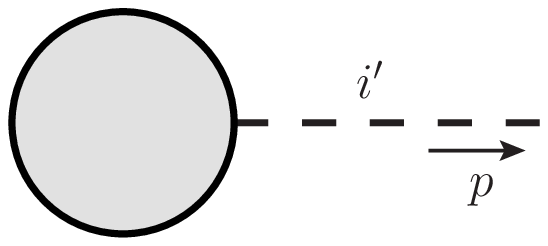}}}.
\label{eq_softexternal}
\end{align}

Summing up all configurations where a gluon connects two external legs and self-energy corrections to each external line, we obtain the complete soft singularities for one-loop amplitude:
\begin{align}
\tilde{\mathcal{M}}^{\rm loop}_{c_1\cdots c_n}\overset{\rm soft}{\sim}&-\frac{1}{2}g_s^2\sum_{i,j}^n\mu^{4-D}\int\limits_{|\vec{k}|<\delta_0} \frac{d^{D-1}k}{2k_0(2\pi)^{D-1}}\frac{p_i\cdot \Pi(k)\cdot p_j}{(k\cdot p_i)(k\cdot p_j)}\bigg|_{k_0=|\vec{k}|}\nonumber \\
&\times \left(\pmb{T}\cdot \pmb{T}\mathcal{M}^{\rm tree}\right)_{c_1\cdots c_i\cdots c_j\cdots c_n} ,
\label{eq_softterm}
\end{align}
where
\begin{equation}
\left(\pmb{T}\cdot \pmb{T}\mathcal{M}^{\rm tree}\right)_{c_1\cdots c_i\cdots c_j\cdots c_n}=
\begin{cases}
T^a_{c_i c_{i^\prime}}T^a_{c_j c_{j^\prime}}\mathcal{M}^{\rm tree}_{c_1\cdots c_{i^\prime}\cdots c_{j^\prime}\cdots c_n},& i\ne j\\
T^a_{c_i c_{i^{\prime\prime}}}T^a_{c_{i^{\prime\prime}} c_{i^\prime}}\mathcal{M}^{\rm tree}_{c_1\cdots c_{i^\prime}\cdots c_n},&i=j
\end{cases}
\end{equation}
is the color connected Born amplitude.

The collinear singularities arise when the virtual gluon is collinear to any massless external momentum.
Here we work in lightcone coordinates where loop momentum $k$ and external momentum $p$ can be expressed as
\begin{align}
&k=(k^+,k^-,\vec{k}_T),\nonumber \\
&p=(p^+,0,\vec{0}_T).
\end{align}
Since the soft-collinear singularities have been incorporated in Eq. \eqref{eq_softterm}, to avoid double counting, we consider only the hard-collinear singularities.
The corresponding scaling behavior of $k$ is $|k^+|\sim \kappa^0$, $|k^-|\sim \kappa^2$ and $|\vec{k}_T|\sim \kappa$.
Then the involved propagators $1/k^2$ and $1/(k\pm p)^2$ scale like $\kappa^{-2}$, other propagators with non-collinear external momenta scale like $\kappa^0$.
The hard-collinear integral region for $k$ is defined as
\begin{equation}
|k^+|>\frac{p^+}{p^0}\delta_0, \quad\quad |\vec{k}_T|<\delta_T,\quad\quad |k^-|\sim \frac{|\vec{k}_T|^2}{|k^+|},
\end{equation}
with the cut on $|k^+|$ exclude the soft-collinear region.
The explicit integration range of $|k^-|$ does not concern with the collinear singular terms.
Note, to validate the scaling behavior of $k$, the soft cutoff parameter should be much larger than the collinear cutoff parameter: $\delta_T\ll \delta_0$.

In lightcone gauge, the structures where virtual gluon $k$ connects the external leg $p$ to hard part (or other external  leg)  scale like $\kappa$:
\begin{equation}
\vcenter{\hbox{\includegraphics[scale=0.5]{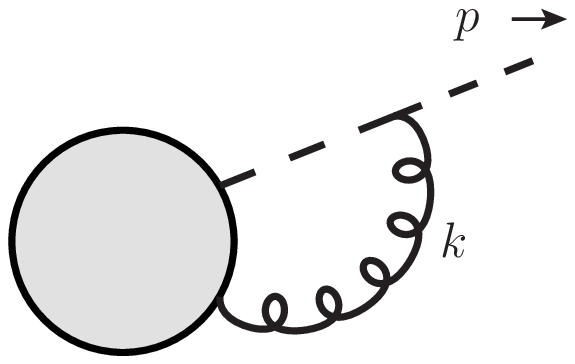}}}\simeq \mathcal{O}(\kappa).
\end{equation}
Therefore, the only collinear singularites come from the self-energy corrections to external legs. We have (see Appendix B for detailed derivation):
\begin{align}
\frac{1}{2}\vcenter{\hbox{\includegraphics[scale=0.5]{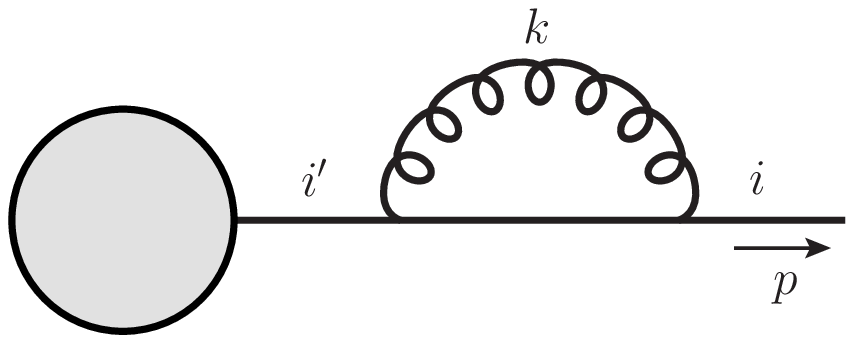}}}\overset{\rm coll}{\sim}&-\frac{g_s^2}{16\pi^2}\frac{1}{\Gamma(1-\epsilon)}\left(\frac{4\pi\mu^2}{\delta_T^2}\right)^\epsilon\frac{1}{\epsilon}C_F\left(\frac{3+\epsilon}{2}+2\ln\frac{\delta_0}{p_0}\right) \nonumber \\
&\times \vcenter{\hbox{\includegraphics[scale=0.5]{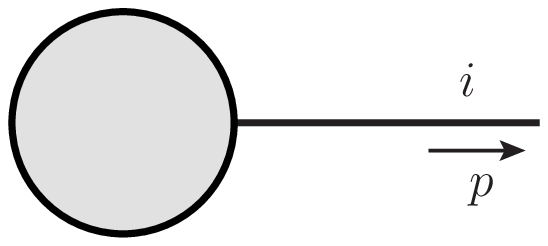}}}, \nonumber \\
\frac{n_{ lf}}{2}\vcenter{\hbox{\includegraphics[scale=0.5]{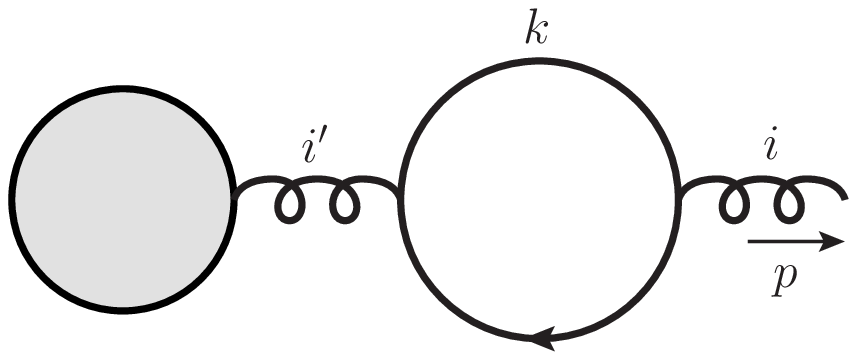}}}\overset{\rm coll}{\sim}&-\frac{g_s^2}{16\pi^2}\frac{1}{\Gamma(1-\epsilon)}\left(\frac{4\pi\mu^2}{\delta_T^2}\right)^\epsilon\frac{1}{\epsilon}n_{lf}\left(-\frac{1-\epsilon}{3-2\epsilon}\right) \nonumber \\
&\times \vcenter{\hbox{\includegraphics[scale=0.5]{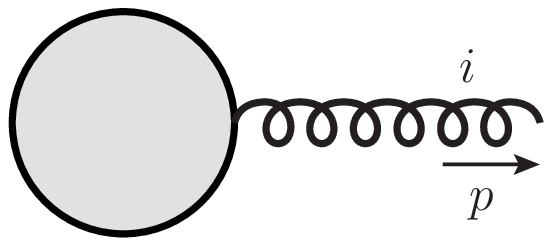}}}, \nonumber \\
\frac{1}{2}\vcenter{\hbox{\includegraphics[scale=0.5]{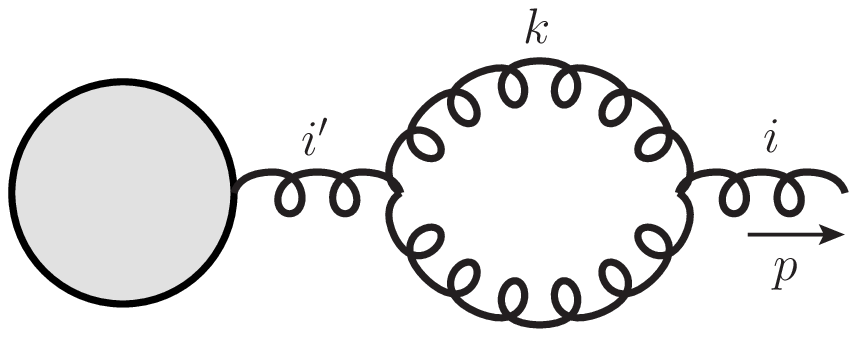}}}\overset{\rm coll}{\sim}&-\frac{g_s^2}{16\pi^2}\frac{1}{\Gamma(1-\epsilon)}\left(\frac{4\pi\mu^2}{\delta_T^2}\right)^\epsilon\frac{1}{\epsilon}C_A\left(\frac{11}{6}+2\ln\frac{\delta_0}{p_0}\right) \nonumber \\
&\times \vcenter{\hbox{\includegraphics[scale=0.5]{colSEGluonB.eps}}}.
\label{eq_collexternal}
\end{align}
Here, $n_{lf}=3$ denotes the number of light quark flavors, $C_A=3$, $C_F=4/3$ are QCD color factors;
the symbol ``$\overset{\rm coll}{\sim}$'' means that the collinear singular terms on each side are equal.
Then the collinear singular terms for one-loop amplitude have the form
\begin{equation}
\tilde{\mathcal{M}}^{\rm loop}_{c_1\cdots c_n}\overset{\rm coll}{\sim}\frac{g_s^2}{16\pi^2}\frac{1}{\Gamma(1-\epsilon)}\left(\frac{4\pi\mu^2}{\delta_T^2}\right)^\epsilon\sum_i\left(\frac{\gamma(i)}{\epsilon}\right)\mathcal{M}^{\rm tree}_{c_1\cdots c_n} ,
\label{eq_collterm}
\end{equation}
where
\begin{align}
&\gamma(q)=-C_F\left(\frac{3+\epsilon}{2}+2\ln\frac{\delta_0}{p_0}\right) ,\nonumber \\
&\gamma(g)=n_{lf}\frac{1-\epsilon}{3-2\epsilon}-C_A\left(\frac{11}{6}+2\ln\frac{\delta_0}{p_0}\right) .
\label{eq_defgamma}
\end{align}

\subsection{Dimensional regularization prescriptions and singular terms in real corrections}
\label{sec_sinR}

The analysis in the previous subsection is based on dimensional regularization.
The key ingredient of dimensional regularization is to continue the dimensions of loop momentum from $4$ to $D=4-2\epsilon$.
While for the treatments of external momenta and gluons' polarization, one is left with some freedom, which result in different variants of dimensional regularization.
Two commonly used variants are\footnote{Another commonly used regularization scheme is the dimensional reduction \cite{Siegel:1979wq,Bern:1991aq,Stockinger:2005gx}, where a quasi-4-dimensional space should be introduced \cite{Stockinger:2005gx}.
The transition rules between dimensional reduction and dimensional regularization are discussed in Refs. \cite{Kunszt:1994np,Catani:1996pk,Signer:2008va,Catani:2000ef}.}:
\begin{itemize}
\item `t Hooft-Veltman (HV) \cite{tHooft:1972tcz} scheme: Loop momentum are treated as $D$-dimensional, while external ones are treated as $4$-dimensional. The gluons inside loop have $D-2$ polarization states, while other gluons have 2 polarizations.
\item Conventional dimensional regularization (CDR) scheme: All momenta are treated as $D$-dimensional, and all gluons have $D-2$ polarization states.
\end{itemize}
The analysis in the previous subsection is legitimate in both HV and CDR schemes.
In Eqs. (\ref{eq_softterm}) and (\ref{eq_collterm}), the only quantity that concerning the choice of HV or CDR is the tree-level amplitude $\mathcal{M}^{\rm tree}$.

\begin{figure}[!thbp]
    \centering
   \includegraphics[scale=0.5]{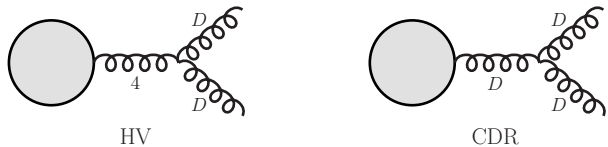}
    \caption{Gluon splitting into soft or collinear gluons under HV and CDR schemes. Here the label $D$ (4) indicates that the momentum of corresponding gluon is $D$- ($4$-) dimensional, and the number of polarization states is $D-2$ (2).}
    \label{fig_realDS}
\end{figure}

In general, only a combination of virtual loop corrections and real emission contributions lead to IR-finite results.
Their dependence on regularization prescriptions should also be canceled as the IR singularities.
This cancellation is only achieved if the regularization prescriptions employed in the virtual and real corrections are consistent, which means unitarity.
Hence, in real corrections, the emitted soft or collinear particles should be treated in the same way as the particles inside loop in virtual corrections.
As an example, Fig. \ref{fig_realDS} shows the case of a gluon splitting into soft or collinear gluons under HV and CDR schemes separately.

There are essentially two types of approaches to evaluate the cross sections of real emission processes: one based on the phase-space slicing method \cite{Fabricius:1981sx,Kramer:1986mc,Harris:2001sx}, and the other based on the subtraction method \cite{Ellis:1980wv,Catani:1996vz,Phaf:2001gc}.
In both approaches, IR singular terms are isolated, and the remaining finite parts can be calculated numerically in 4-dimensional space-time.
To match our analysis on virtual corrections, here we take the two cutoff phase-space slicing method.
As the corresponding implementation is described in detail in Ref. \cite{Harris:2001sx}, we only introduce some main results here.

Considering the real emission process
\begin{equation}
p_a+p_b\to p_1+\cdots +p_n+p_{n+1}\ ,
\end{equation}
where $(n+1)$ is the ``additional'' particle that may soft or collinear to another massless external line.
By introducing soft cut $\delta_0$ and collinear cut $\delta_T$\footnote{The cutoff parameters used here is different from that used in Ref. \cite{Harris:2001sx}. They are related by the relations $\delta_0=\frac{\sqrt{s_{12}}}{2}\delta_s^\text{\tiny\cite{Harris:2001sx}}$ and $\delta_T^2=z(1-z)\delta_c^\text{\tiny\cite{Harris:2001sx}}s_{12}$.} which satisfies $\delta_0\gg \delta_T$, the phase space can be separated into three regions:
\begin{itemize}
\item soft: $p_{n+1}^0<\delta_0$;
\item hard-collinear (HC): $p_{n+1}^0>\delta_0$ and $|\vec{p}_{n+1\ T}|<\delta_T$;
\item hard-non-collinear (HNC): $p_{n+1}^0>\delta_0$ and $|\vec{p}_{n+1\ T}|>\delta_T$.
\end{itemize}
Here, the transverse momentum $|\vec{p}_{n+1\ T}|$ is relative to the ``parent'' particle $i^\prime$, which splitting into $i$ and $(n+1)$: $i^\prime\to i+(n+1)$.
The cross section of real emission process can be written as
\begin{equation}
\sigma_{\rm real}=\sigma_{\rm real}^{\rm soft}+\sigma_{\rm real}^{\rm HC}+\sigma_{\rm real}^{\rm HNC}.
\end{equation}

After neglecting terms of order $\delta_0$ and $\delta_T$, the soft and hard-collinear pieces take the form\footnote{Here we present the results for indistinguishable final state case only. The results for other cases, like tagged final state or hadron in initial state, can be found in Ref. \cite{Harris:2001sx}.}
\begin{align}
\sigma_{\rm real}^{\rm soft}=&\frac{1}{2\Phi}\int\limits_{\rm S}d\Gamma_{n+1}\overline{\sum}|\mathcal{M}^{\rm real}_{c_1\cdots c_{n+1}}|^2\nonumber \\
=&g_s^2\mu^{4-D} \int\limits_{|\vec{p}_{n+1}|<\delta_0}\frac{d^{D-1}p_{n+1}}{2p_{n+1}^0(2\pi)^{D-1}}\sum_{i,j}^n\bigg\{\frac{p_i\cdot \Pi(p_{n+1})\cdot p_j}{(p_i\cdot p_{n+1})(p_j\cdot p_{n+1})}\nonumber \\
&\frac{1}{2\Phi}\int d\Gamma_n\overline{\sum}\left[\mathcal{M}^{\rm tree}_{c_1\cdots c_{i^\prime}\cdots c_j\cdots c_n}T^a_{c_ic_{i^\prime}}\right]\left[\mathcal{M}^{\rm tree}_{c_1\cdots c_i\cdots c_{j^\prime}\cdots c_n}T^a_{c_jc_{j^\prime}}\right]^*\bigg\} ,
\label{eq_realsoft}
\end{align}
and
\begin{align}
\sigma_{\rm real}^{\rm HC}=&\frac{1}{2\Phi}\int\limits_{\rm HC}d\Gamma_{n+1}\overline{\sum}|\mathcal{M}^{\rm real}_{c_1\cdots c_{n+1}}|^2\nonumber \\
=&-\frac{g_s^2}{8\pi^2}\frac{1}{\Gamma(1-\epsilon)}\left(\frac{4\pi\mu^2}{\delta_T^2}\right)^\epsilon\left(\sum_i\frac{\gamma(i)}{\epsilon}\right) \frac{1}{2\Phi}\int d\Gamma_n\overline{\sum}|\mathcal{M}^{\rm tree}_{c_1\cdots c_n}|^2 ,
\label{eq_realcollinear}
\end{align}
where $\Phi$ is the flux factor, $d\Gamma_n$ and $d\Gamma_{n+1}$ stand for the $n$- and $(n+1)$- body phase space respectively, $\gamma(i)$ has been defined in Eq. \eqref{eq_defgamma}.

Comparing Eqs. \eqref{eq_realsoft} and \eqref{eq_realcollinear} with Eqs. \eqref{eq_softterm} and \eqref{eq_collterm}, we can see that the IR singularities in real corrections are canceled explicitly against their counterparts in virtual corrections, as expected by the Kinoshita-Lee-Nauenberg (KLN) theorem \cite{Kinoshita:1962ur,Lee:1964is}.
Their dependence on dimensional regularization prescription are also canceled, as long as the corresponding $\gamma(i)$ and $\mathcal{M}^{\rm tree}$ are obtained with the same scheme.

\section{Generalize the TAM to NLO QCD calculation}
\label{sec_TATNLO}

In this section, we discuss how to apply TAM in NLO QCD calculations.
We also compare the result obtained from TAM with that from CAS.

\subsection{Calculation scheme}
\label{sec_TATNLOScheme}

In fact, the application of TAM in the calculation of one-loop helicity amplitude is transparent, and will not lead to any additional technical difficulty.
The only concern is that one should find a proper scheme for real corrections to guarantee the unitarity.

In Section \ref{sec_sinV}, we analyze the singularity structure of one-loop amplitude.
The analysis there is performed at amplitude level, and is irrelevant with the treatment of spinors.
Therefore, for one-loop helicity amplitude which calculated through TAM, the formulas (\ref{eq_UVterm}), (\ref{eq_softterm}) and (\ref{eq_collterm}) still hold.
While in real corrections, the singularity formulas (\ref{eq_realsoft}) and (\ref{eq_realcollinear}) are at cross section (decay width) level. We can decompose $\sigma_{\rm real}^{\rm soft}$ and $\sigma_{\rm real}^{\rm HC}$ according to the helicity states of the $1$st to $n$-th particles, which leads to
\begin{align}
&\sum_{\lambda_{n+1}}\sigma_{\rm real}^{\rm soft}(\lambda_1,\cdots,\lambda_n,\lambda_{n+1})\nonumber \\
=&g_s^2\mu^{4-D} \int\limits_{|\vec{p}_{n+1}|<\delta_0}\frac{d^{D-1}p_{n+1}}{2p_{n+1}^0(2\pi)^{D-1}}\sum_{i,j}^n\bigg\{\frac{p_i\cdot \Pi(p_{n+1})\cdot p_j}{(p_i\cdot p_{n+1})(p_j\cdot p_{n+1})}\nonumber \\
&\frac{1}{2\Phi}\int d\Gamma_n\left[\mathcal{M}^{\rm tree}_{c_1\cdots c_{i^\prime}\cdots c_j\cdots c_n}(\lambda_1,\cdots,\lambda_n)T^a_{c_ic_{i^\prime}}\right]\left[\mathcal{M}^{\rm tree}_{c_1\cdots c_i\cdots c_{j^\prime}\cdots c_n}(\lambda_1,\cdots,\lambda_n)T^a_{c_jc_{j^\prime}}\right]^*\bigg\},
\label{eq_realsoftH}
\end{align}
and
\begin{align}
&\sum_{\lambda_{n+1}}\sigma_{\rm real}^{\rm HC}(\lambda_1,\cdots,\lambda_n,\lambda_{n+1})\nonumber \\
=&-\frac{g_s^2}{8\pi^2}\frac{1}{\Gamma(1-\epsilon)}\left(\frac{4\pi\mu^2}{\delta_T^2}\right)^\epsilon\left(\sum_i\frac{\gamma(i)}{\epsilon}\right) \frac{1}{2\Phi}\int d\Gamma_n|\mathcal{M}^{\rm tree}_{c_1\cdots c_n}(\lambda_1,\cdots,\lambda_n)|^2 \ .
\label{eq_realcollinearH}
\end{align}
Here, $\lambda_i$ denotes the helicity state of the $i$-th particle, $\mathcal{M}^{\rm tree}(\lambda_1,\cdots,\lambda_n)$ denotes the tree-level helicity amplitude, which should be calculated through the TAM to guarantee the unitarity.

With the above preparation, we then obtain the scheme that can be used in NLO QCD calculations: the one-loop helicity amplitude can be calculated straightforwardly by using the TAM,
 and the soft and hard-collinear pieces of real corrections should be calculated through Eqs. (\ref{eq_realsoftH}) and (\ref{eq_realcollinearH}).
There are some remarkable points in practice as follows:

1) As we have discussed in Sec. \ref{sec_sinR}, this calculation scheme is legitimate under both CDR and HV schemes.

2) In the TAM, additional $\gamma_5$ is introduced in Eq. \eqref{eq_lamga5}.
Although in dimensional regularization, $\gamma_5$ is always a difficult object to deal with, the $\gamma_5$ here will not cause additional trouble. Because this $\gamma_5$ is only concerned with the treatment of spinor, which is independent with singularity structure.
Hence, our scheme is valid under arbitrary self-consistent $\gamma_5$ prescription, like the `t Hooft-Veltman-Breitenlohner-Maison (HVBM) prescription \cite{tHooft:1972tcz,Breitenlohner:1976te} or the Kreimer-Korner prescription \cite{Kreimer:1989ke,Korner:1991sx}.

3) The integrand in Eq. \eqref{eq_realsoftH} contains the factor
\begin{equation}
\frac{p^{}_i\cdot \Pi(p^{}_{n+1})\cdot p^{}_j}{(p^{}_i\cdot p^{}_{n+1})(p^{}_j\cdot p^{}_{n+1})}=\frac{-p^{}_i\cdot p^{}_j}{(p^{}_i\cdot p^{}_{n+1})(p^{}_j\cdot p^{}_{n+1})} +\frac{1}{r\cdot p^{}_{n+1}}\left(\frac{r\cdot p^{}_i}{p^{}_i\cdot p^{}_{n+1}}+\frac{r\cdot p^{}_j}{p^{}_j\cdot p^{}_{n+1}}  \right).
\label{eq_lcnum}
\end{equation}
By using the color conservation relation $\sum_{i=1}^nT^a_{c_ic_{i^\prime}}\mathcal{M}^{\rm tree}_{c_1\cdots c_{i^\prime}\cdots c_n}=0$ \cite{Catani:2000ef}, we can see that the second term is vanished after summing over $i$ and $j$.
Hence in actual calculation, we can replace $p^{}_i\cdot \Pi(p^{}_{n+1})\cdot p^{}_j$ by $-p^{}_i\cdot p^{}_j$.

\subsection{Consistency between TAM and CAS}
\label{sec_TandC}

To compare the result obtained from TAM with that from CAS, we introduce a quantity $\mathcal{F}(\Gamma_A,\Gamma_B)$:
\begin{equation}
\mathcal{F}(\Gamma_A,\Gamma_B)=\sum_{\lambda_1,\lambda_2}\bar{U}(p_2,\lambda_2)\Gamma_AU(p_1,\lambda_1)\bar{U}(p_1,\lambda_1)\Gamma_BU(p_2,\lambda_2)\ ,
\end{equation}
where $\Gamma_A$ and $\Gamma_B$ denote series of Dirac gamma matrices.
This quantity can be calculated through either CAS or TAM:
\begin{align}
\mathcal{F}_{\rm CAS}(\Gamma_A,\Gamma_B)=&{\rm tr}[\Gamma_A(\slashed{p}_1+M_1)\Gamma_B(\slashed{p}_2+M_2)]\ ,\nonumber \\
\mathcal{F}_{\rm TAM}(\Gamma_A,\Gamma_B)=&\frac{1}{16(k_0\cdot p_1)(k_0\cdot p_2)}\sum_{\lambda_1,\lambda_2}\Big\{{\rm tr}[\Gamma_A(\slashed{p}_1+M_1)\Lambda(\lambda_1,\lambda_2)(\slashed{p}_2+M_2) ] \nonumber \\
&{\rm tr}[\Gamma_B(\slashed{p}_2+M_2)\Lambda(\lambda_2,\lambda_1)(\slashed{p}_1+M_1)]\Big\}\ .
\end{align}
In 4-dimensions, the gamma matrices can be represented explicitly by $4\times 4$ matrices, and the construction (\ref{eq_construct}) is compatible with the fermion spin sum relation (\ref{eq_fspinsum}), which means $\mathcal{F}_{\rm TAM}(\Gamma_A,\Gamma_B)$ is equivalent with $\mathcal{F}_{\rm CAS}(\Gamma_A,\Gamma_B)$.
While in general $D$-dimensions, this equivalence will no longer hold\footnote{For example, considering $\Gamma_A=\gamma^\mu$, $\Gamma_B=\gamma_\mu$. By using the HVBM $\gamma_5$-scheme, we obtain $\mathcal{F}_{\rm CAS}(\gamma^\mu,\gamma_\mu)=4DM_1M_2-(4D-8)p_1\cdot p_2$, and $\mathcal{F}_{\rm TAM}(\gamma^\mu,\gamma_\mu)=16M_1M_2-8p_1\cdot p_2$.}.
The subtle difficulty is that there is inconsistency between the continuous space-time dimensions and the fixed spinor space dimensions, as revealed in Ref. \cite{Siegel:1980qs}.
Therefore, we may write
 \begin{equation}
\mathcal{F}_{\rm CAS}(\Gamma_A,\Gamma_B)-\mathcal{F}_{\rm TAM}(\Gamma_A,\Gamma_B)=\mathcal{O}(\epsilon).
\label{eq_diff2app}
\end{equation}
At NLO,  one may worry that the $\mathcal{O}(\epsilon)$ term will interfere with UV or IR singularities, which eventually lead to terms that are finite or even divergent as $\epsilon\to 0$.
In fact, the $\mathcal{O}(\epsilon)$ term arises due to different treatment of spinor, whose effect will be eliminated as we sum up all pieces of NLO corrections, as discussed in the preceding subsection.

In HV dimensional regularization scheme, all momenta and gluons outside loop are treated as 4-dimensional.
The product of one-loop amplitude and Born amplitude takes the form $\mathcal{F}(\Gamma_A,\overline{\Gamma}_B)$, where $\overline{\Gamma}_B$ denotes a series of 4-dimensional Dirac gamma matrices that comes from Born amplitude. In this case, we have
 \begin{equation}
\mathcal{F}_{\rm CAS}(\Gamma_A,\overline{\Gamma}_B)-\mathcal{F}_{\rm TAM}(\Gamma_A,\overline{\Gamma}_B)=0,
\end{equation}
which indicates that the consistency can be obtained even without a combination of virtual and real corrections.

\section{Example}
To demonstrate the calculation scheme proposed in Sec. \ref{sec_TATNLO}, in this section, we apply it to the calculation of the NLO QCD corrections to $g+g\to t+\bar{t}$ and $q+\bar{q}\to t+\bar{t}$ processes.
We find that the final results are consistent with that from the CAS.
Another successful application of the TAM at one-loop level can be found in Ref. \cite{Chen:2012ju}, where Higgs boson decays to $l\bar{l}\gamma$ was studied.

The momenta and polarization states of incoming and outgoing particles are denoted as:
\begin{align}
&g(p_1,\lambda_1)+g(p_2,\lambda_2)\to t(p_3,\lambda_3)+\bar{t}(p_4,\lambda_4), \nonumber \\
&q(p_1,\lambda_1)+\bar{q}(p_2,\lambda_2)\to t(p_3,\lambda_3)+\bar{t}(p_4,\lambda_4) .
\end{align}
Here, initial and final state particles are all on their mass shells: $p_1^2=p_2^2=0$ and $p_3^2=p_4^2=m^{2}_{t}$.
The gluons' polarization vectors are denoted as $\epsilon_1^{(\lambda_1)}$ and $\epsilon_2^{(\lambda_2)}$, which satisfy the constrains $\epsilon_1^{(\lambda_1)}\cdot \epsilon_1^{(\lambda_1)*}=\epsilon_2^{(\lambda_2)}\cdot \epsilon_2^{(\lambda_2)*}=-1$ and $p_1\cdot \epsilon_1^{(\lambda_1)}=p_2\cdot \epsilon_2^{(\lambda_2)}=0$.

In the center-of-mass system,  the momenta and gluons' polarization vectors are chosen as:
\begin{align}
&p_1=\frac{\sqrt{s}}{2}(1,0,0,1) , \quad p_2=\frac{\sqrt{s}}{2}(1,0,0,-1) , \nonumber \\
&p_3=\frac{\sqrt{s}}{2}(1,0,r_y,r_z) , \quad p_4=\frac{\sqrt{s}}{2}(1,0,-r_y,-r_z) ,
\label{eq_momchoice}
\end{align}
and
\begin{equation}
\epsilon_1^{(1)}=\epsilon_2^{(1)}=(0,1,0,0) ,\quad \epsilon_1^{(2)}=\epsilon_2^{(2)}=(0,0,1,0) ,
\end{equation}
where $s=(p_1+p_2)^2$, and the on-shell condition constrains $r_y^2+r_z^2=1-4m_t^2/s$.
The helicity states of fermions are defined under the KS basis with the auxiliary vectors chosen as:
\begin{equation}
 k_0=(1,1,0,0) ,\quad k_1=(0,0,0,1) .
\end{equation}
Then the tree-level and one-loop helicity amplitudes can be calculated straightforwardly by using the spinor product formula (\ref{eq_sProS}).

In the computation of one-loop amplitudes, the HV dimensional regularization is adopted to regularize the UV and IR singularities.
For the $\gamma_5$ introduced in Eq. \eqref{eq_sProS}, the HVBM prescription is adopted.
The UV singularities are removed by renormalization procedure.
The renormalization constants include $Z_2$, $Z_m$, $Z_l$, $Z_3$ and $Z_g$, corresponding to heavy quark field, heavy quark mass, light quark field, gluon field and strong coupling constant, respectively.
We define $Z_2$, $Z_m$, $Z_l$ and $Z_3$ in the on-shell (OS) scheme, $Z_g$ in the modified minimal-subtraction ($\overline{\rm MS}$) scheme. The corresponding counterterms are
\begin{align}
\delta Z_2^{\rm OS}=&-C_F\frac{\alpha_s}{4\pi}\left[\frac{1}{\epsilon_{\rm UV}}+\frac{2}{\epsilon_{\rm IR}}
-3\gamma_E+3\ln\frac{4\pi\mu^2}{m_t^2}+4\right] ,\nonumber\\
\delta Z_m^{\rm OS}=&-3C_F\frac{\alpha_s}{4\pi}\left[\frac{1}{\epsilon_{\rm UV}}-\gamma_E+\ln\frac{4\pi\mu^2}{m_t^2} +\frac{4}{3}\right] , \nonumber\\
 \delta Z_l^{\overline{\rm OS}}=&-C_F \frac{\alpha_s}{4\pi}
  \left[\frac{1}{\epsilon_{\rm UV}} - \frac{1}{\epsilon_{\rm IR}} \right] , \nonumber \\
\delta Z_3^{\overline{\rm OS}}=&\frac{\alpha_s}{4\pi}\left[(\beta^\prime_0-2C_A)\left(\frac{1}{\epsilon_{\rm UV}} -\frac{1}{\epsilon_{\rm IR}}\right)-\frac{4}{3}T_F\sum_{i=c,b,t}\left(\frac{1}{\epsilon_{\rm UV}}-\gamma_E+\ln\frac{4\pi\mu^2}{m_i^2}\right) \right] , \nonumber\\
 \delta Z_g^{\overline{\rm MS}}=&-\frac{\beta_0}{2}\frac{\alpha_s}{4\pi}\left[\frac{1}{\epsilon_{\rm UV}} -\gamma_E + \ln(4\pi)  \right].
\end{align}
Here, $\mu$ is the renormalization scale, $\gamma_E$ is the Euler's constant;
$C_A=3$, $C_F=4/3$ and $T_F=1/2$ are QCD color factors;
$\beta_0=(11/3)C_A-(4/3)T_Fn_f$ is the one-loop coefficient of QCD beta function, in which $n_f=6$ is the number of active quark flavors,
$\beta^\prime_0=(11/3)C_A-(4/3)T_Fn_{lf}$ and $n_{lf}=3$ is the number of light quark flavors.

In real corrections, IR singularities arise from the phase-space integration of the additional emitted gluon, whose momentum is denoted by $p_5$ hereafter.
To isolate the singularities, we follows the lines of Ref. \cite{Harris:2001sx}.
By introducing two cutoff parameters $\delta_s$ and $\delta_c$, the real correction phase space is split into three regions:
\begin{itemize}
\item soft: $p_5^0\le\delta_s\sqrt{s}/2$;
\item hard-collinear: including the collinear-to-$p_1$ region where $p_5^0>\delta_s\sqrt{s}/2$ and $(p_1+p_5)^2\le \delta_c s$, and the  the collinear-to-$p_2$ region where $p_5^0>\delta_s\sqrt{s}/2$ and $(p_2+p_5)^2\le \delta_c s$.
\item hard-non-collinear: $p_5^0>\delta_s\sqrt{s}/2$ and $(p_1+p_5)^2> \delta_c s$ and $(p_2+p_5)^2> \delta_c s$.
\end{itemize}
The soft and hard-collinear contributions can be obtained through Eqs. (2.22) and (2.74) of Ref. \cite{Harris:2001sx}, 
with the unpolarized Born cross section $\sigma^0$ replaced by helicity cross section $\sigma^0(\lambda_1,\lambda_2,\lambda_3,\lambda_4)$, which should be calculated through the TAM.
The remaining hard-non-collinear contribution is IR finite, and can be calculated numerically in 4-dimensions.
After summing up these three pieces, their dependence on technical cuts are eliminated as expected.

The total NLO corrections are obtained by summing up the virtual and real corrections.
We verify that all singularities are canceled exactly.
By taking the same renormalization scheme, we find that our results agree with Fig. 3 and Fig. 4 of Ref. \cite{Nason:1987xz} within 1\% accuracy.

For illustrative purpose, we present the numerical values of $\mathcal{A}^{\rm loop}$, $\mathcal{A}^{\rm CT}$, $\mathcal{A}^{\rm soft}_{\rm real}$ and $\mathcal{A}^{\rm HC}_{\rm real}$, which are defined as:
\begin{align}
&\mathcal{A}^{\rm loop}=2\sum_{\lambda}{\rm Re}\left[\mathcal{M}^{\rm loop}(\mathcal{M}^{\rm tree})^*\right],\nonumber \\
&\mathcal{A}^{\rm CT}=2\sum_{\lambda}{\rm Re}\left[\mathcal{M}^{\rm CT}(\mathcal{M}^{\rm tree})^*\right],\nonumber \\
&\mathcal{A}^{\rm soft}_{\rm real}=g_s^2\mu^{2\epsilon}\int\limits_{p_5^0<\delta_s\sqrt{s}/2}\frac{d^{D-1}p_5}{2p_5^0(2\pi)^{D-1}}\sum_{i,j=1}^4\frac{p_i\cdot p_j}{(p_i\cdot p_5)(p_j\cdot p_5)}\nonumber \\
&\quad\quad\quad \sum_{\lambda}\left[\mathcal{M}^{\rm tree}_{c_1\cdots c_{i^\prime}\cdots c_j\cdots c_4}T^a_{c_ic_{i^\prime}}\right]\left[\mathcal{M}^{\rm tree}_{c_1\cdots c_i\cdots c_{j^\prime}\cdots c_4}T^a_{c_jc_{j^\prime}}\right]^*,\nonumber \\
&\mathcal{A}^{\rm HC}_{\rm real}=\frac{g_s^2}{4\pi^2}\frac{1}{\Gamma(1-\epsilon)}\left(\frac{4\pi\mu^2} {\mu_f^2}\right)\frac{A_1}{\epsilon}\sum_{\lambda}|\mathcal{M}^{\rm tree}|^2 \ .
\end{align}
Here, $\mu_f$ is the initial state factorization scale, and for gluon-gluon channel $A_1=-n_{lf}/3+C_A(11/6+\ln\delta_s)$ , for quark-antiquark channel $A_1=C_F(3/2+2\ln\delta_s)$.
Note, for comparison convenience, the hard-collinear contribution $\mathcal{A}^{\rm HC}_{\rm real}$ here only  include singular or approach dependent (TAM or CAS) terms.

The mass of heavy quarks are taken as $m_t=173$, $m_b=4.9$, $m_c=1.5$, and the strong coupling $g_s$ is set to one.
The numerical results for $g+g\to t+\bar{t}$ and $q+\bar{q}\to t+\bar{t}$ processes at the point $s=500^2$, $r_z=0.6$, $\mu_r=\mu_f=500$ are given in Table \ref{tab_gg} and Table \ref{tab_qq}, respectively.
As a comparison, we also list the results obtained from the CAS, where CDR scheme is employed.
It can be seen that for each piece of NLO corrections, the results from TAM and CAS are generally different.
However, after summing up all pieces, consistent results are obtained.
The numerical results confirm our statement in Sec. \ref{sec_TandC}.

\begin{table}[!htbp]
\caption{Numerical results for $g+g\to t+\bar{t}$ process.}
\centering
  \begin{tabular}{|p{1cm}<{\centering}|p{7cm}<{\centering}|p{7cm}<{\centering}|}
 \hline
\,    & TAM with HV & CAS with CDR  \\
    \hline
       $\mathcal{A}^{\rm loop}$ &
        \begin{tabular}{@{}c@{}} $\scriptstyle -10.08118054436082\epsilon ^{-2}-3.437853927720160\epsilon^{-1}$\\ $\scriptstyle +75.75337372837709$\end{tabular}
         &
         \begin{tabular}{@{}c@{}} $\scriptstyle -10.08118054436082\epsilon ^{-2}+17.83506953428062\epsilon^{-1}$\\ $\scriptstyle +69.74008746354726$\end{tabular}
         \\
     \hline
      $\mathcal{A}^{\rm CT}$ & $\scriptstyle -20.99049132259232\epsilon^{-1}-87.04718302407158$ &
      $\scriptstyle -20.47530745414479\epsilon^{-1}-41.20686937309857$
      \\
 \hline
       $\mathcal{A}^{\rm soft}_{\rm real}$ &
       \begin{tabular}{@{}c@{}} $\scriptstyle 10.08118054436082\epsilon ^{-2}+9.306574433771248\epsilon^{-1}$\\ $\scriptstyle -20.16236108872165 \ln \delta_s \epsilon^{-1}-20.38707598960415$\\$\scriptstyle +20.16236108872165 \ln ^2\delta_s-18.61314886754250 \ln \delta_s$\end{tabular}
        &
        \begin{tabular}{@{}c@{}} $\scriptstyle 10.08118054436082\epsilon ^{-2}-12.48153289667707\epsilon^{-1}$\\ $\scriptstyle -20.16236108872165 \ln \delta_s \epsilon^{-1}-27.53194238007485$\\$\scriptstyle +20.16236108872165 \ln ^2\delta_s+24.96306579335413 \ln \delta_s$\end{tabular}
        \\
 \hline
       $\mathcal{A}^{\rm HC}_{\rm real}$ &
       \begin{tabular}{@{}c@{}} $\scriptstyle 20.16236108872165\ln\delta_s \epsilon^{-1}+15.12177081654124 \epsilon^{-1} $ \\ $\scriptstyle -39.39339412989338\ln \delta_s+29.54504559742003$ \end{tabular}
        &
         \begin{tabular}{@{}c@{}} $\scriptstyle 20.16236108872165\ln\delta_s \epsilon^{-1}+15.12177081654124 \epsilon^{-1} $ \\ $\scriptstyle -4.182820531003251\ln \delta_s-3.137115398252439$ \end{tabular}
        \\
 \hline
 sum &
 \begin{tabular}{@{}c@{}} $\scriptstyle 20.78024526235088 \ln\delta_s+20.16236108872165 \ln^2\delta_s$ \\ $\scriptstyle -2.1358396878786$ \end{tabular}
 &
  \begin{tabular}{@{}c@{}} $\scriptstyle 20.78024526235088 \ln\delta_s+20.16236108872165 \ln^2\delta_s$ \\ $\scriptstyle -2.1358396878786$ \end{tabular}
 \\
 \hline
  \end{tabular}
\label{tab_gg}
\end{table}

\begin{table}[!htbp]
\caption{Numerical results for $q+\bar{q}\to t+\bar{t}$ process.}
\centering
  \begin{tabular}{|p{1cm}<{\centering}|p{7cm}<{\centering}|p{7cm}<{\centering}|}
 \hline
\,    & TAM with HV & CAS with CDR  \\
    \hline
       $\mathcal{A}^{\rm loop}$ &
        \begin{tabular}{@{}c@{}} $\scriptstyle -0.3998042041986813\epsilon ^{-2}+0.6577945202719476\epsilon^{-1}$\\ $\scriptstyle +6.474889692802957$\end{tabular}
         &
         \begin{tabular}{@{}c@{}} $\scriptstyle -0.3998042041986813\epsilon ^{-2}+1.198174166364416\epsilon^{-1}$\\ $\scriptstyle +5.585807571271599$\end{tabular}
         \\
     \hline
      $\mathcal{A}^{\rm CT}$ & $\scriptstyle -1.649192342319560\epsilon^{-1}-5.294770961065026$ &
      $\scriptstyle -1.649192342319560\epsilon^{-1}-3.065704920933595$
      \\
 \hline
       $\mathcal{A}^{\rm soft}_{\rm real}$ &
       \begin{tabular}{@{}c@{}} $\scriptstyle 0.399804204198681\epsilon ^{-2}+0.3916915157495908\epsilon^{-1}$\\ $\scriptstyle-0.7996084083973625 \ln \delta_s \epsilon^{-1}-1.043296995492568$\\$\scriptstyle +0.7996084083973625 \ln ^2\delta_s-0.7833830314991816 \ln \delta_s$\end{tabular}
        &
        \begin{tabular}{@{}c@{}} $\scriptstyle 0.399804204198681\epsilon ^{-2}-0.1486881303428773\epsilon^{-1}$\\ $\scriptstyle -0.7996084083973625 \ln \delta_s \epsilon^{-1}-1.572711444953939$\\$\scriptstyle +0.7996084083973625 \ln ^2\delta_s+0.2973762606857546 \ln \delta_s$\end{tabular}
        \\
 \hline
       $\mathcal{A}^{\rm HC}_{\rm real}$ &
       \begin{tabular}{@{}c@{}} $\scriptstyle 0.7996084083973625\ln\delta_s \epsilon^{-1}+0.5997063062980219 \epsilon^{-1} $ \\ $\scriptstyle1.562281770620308\ln \delta_s+1.171711327965231$ \end{tabular}
        &
         \begin{tabular}{@{}c@{}} $\scriptstyle 0.7996084083973625\ln\delta_s \epsilon^{-1}+0.5997063062980219 \epsilon^{-1} $ \\ $\scriptstyle +0.4815224784353714\ln \delta_s+0.3611418588265285$ \end{tabular}
        \\
 \hline
 sum &
 \begin{tabular}{@{}c@{}} $\scriptstyle 0.7788987391211260 \ln\delta_s+0.7996084083973625 \ln^2\delta_s$ \\ $\scriptstyle +1.30853306421059$ \end{tabular}
 &
  \begin{tabular}{@{}c@{}} $\scriptstyle 0.7788987391211260 \ln\delta_s+0.7996084083973625 \ln^2\delta_s$ \\ $\scriptstyle +1.30853306421059$ \end{tabular}
 \\
 \hline
  \end{tabular}
\label{tab_qq}
\end{table}

\section{Summary}

Helicity amplitude method is not only a technique to perform the perturbative calculation of Feynman diagrams, it provides more information than conventional amplitude squaring approach in phenomenological study. In this paper, we reviewed the basic idea of TAM, and discussed how to generalize this method to NLO QCD calculation.
By analyzing the singularity structures of virtual and real corrections, we proposed a scheme that can guarantee the unitarity in the NLO QCD calculation. This scheme is legitimate under both CDR and HV schemes, and is compatible with arbitrary self-consistent $\gamma_5$  prescription. We also provided an illustrative example by this scheme.
Another noteworthy aspect of this work is that, instead of relying on KLN theorem, we shown the cancellations of soft and collinear singularities explicitly, at least at NLO, by using the power counting technique.

Last, it should be mentioned that the IR divergence cancellation at next-to-next-to-leading order (NNLO) is much more complicated than that at NLO. To tackle this issue, the techniques developed in Refs. \cite{Grammer:1973db,Catani:1998bh,Becher:2009kw,Becher:2009qa,Feige:2014wja} may be useful.

\vspace{0.3cm} {\bf Acknowledgments}

This work was supported in part by the National Natural Science Foundation of China(NSFC) under the Grants 11975236  ,11635009, and 12047553.

\section*{Appendix A: Soft singularities of external self-energy diagrams}

In this appendix, we present the derivation of Eq. \eqref{eq_softexternal}.
As we work in lightcone gauge, a proper prescription should be introduced to treat the unphysical pole $(r\cdot k)^{-1}$.
It has been shown that the usual principal-value prescription is incompatible with Wick rotation, and will lead to a violation of Ward identity \cite{Leibbrandt:1983pj}.
These difficulties can be avoided by using the Mandelstam-Leibbrandt (ML) prescription \cite{Mandelstam:1982cb,Leibbrandt:1983pj}:
\begin{equation}
\frac{1}{k\cdot r}\to \frac{k\cdot r^*}{(k\cdot r^*) (k\cdot r)+i\varepsilon}=\frac{1}{k\cdot r+i\varepsilon\ {\rm sign}(r^*\cdot k)}\ ,
\label{eq_ML}
\end{equation}
where $r^*=(r^0,-\vec{r})$. Although our analysis here does not involve any explicit computation of Feynman integral, we as well use the ML prescription.

The self-energy correction to external massive quark takes the form
\begin{equation}
\frac{1}{2}\vcenter{\hbox{\includegraphics[scale=0.5]{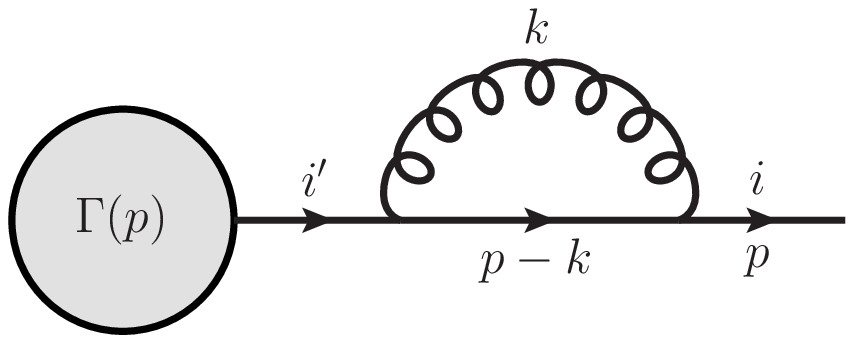}}}=\frac{1}{2}\bar{u}(p)[-i\Sigma_{c_ic_{i^\prime}}(p)]\frac{i(\slashed{p}+m)}{p^2-m^2}\Gamma_{c_{i^\prime}}(p)\bigg|_{p_0\to \omega_p},
\end{equation}
where $\omega_p=\sqrt{|\vec{p}|^2+m^2}$, and
\begin{equation}
-i\Sigma_{c_ic_{i^\prime}}(p)=g_s^2C_F\delta_{c_ic_{i^\prime}}\mu^{4-D}\int\frac{d^Dk}{(2\pi)^D}\frac{\gamma^\mu(\slashed{p}-\slashed{k}+m)\gamma^\nu}{k^2[(p-k)^2-m^2]}\Pi_{\mu\nu}(k).
\label{eq_qselint}
\end{equation}
By performing the Passarino-Veltman tensor reduction, the $-i\Sigma_{c_ic_{i^\prime}}(p)$ can be reduced to the form\footnote{The vector $r^*$ is introduced by the ML prescription. Since our derivation does not involve explicit integral computation, the final expression should be prescription independent. As a check, we performed an analysis without any $r^*$ involved, and consistent result was obtained.}
\begin{equation}
-i\Sigma_{c_ic_{i^\prime}}(p)=g_s^2C_F\delta_{c_ic_{i^\prime}}[f_1m+f_2(\slashed{p}-m)+f_3\slashed{r}+f_4\slashed{r}^*],
\end{equation}
where $f_i$ should be expand near $p_0=\omega_p$:
\begin{equation}
f_i=f_i^{(0)}+f_i^{(1)}(p_0-\omega_p)+\mathcal{O}\left((p_0-\omega_p)^2\right),
\end{equation}
with
\begin{equation}
f_i^{(n)}=\frac{d^nf_i}{dp_0^n}\bigg|_{p_0=\omega_{\vec{p}}} .
\end{equation}
By rescaling the loop momentum as $k\to \kappa^2 k$, we can perform power counting to pick the terms that are potentially soft divergent:
\begin{align}
f_1^{(0)}\overset{\rm soft}{\sim}& 0 ,\quad  f_2^{(0)}\overset{\rm soft}{\sim} 0\ ,\quad  f_3^{(0)}\overset{\rm soft}{\sim} 0\ ,\quad  f_4^{(0)}\overset{\rm soft}{\sim} 0 , \nonumber \\
f_1^{(1)}\overset{\rm soft}{\sim}&\frac{2(p\cdot r)}{m^2(r\cdot r^*)-2(p\cdot r)(p\cdot r^*)}[\omega_p(r\cdot r^*)I_1-(p\cdot r^*)I_2-(p\cdot r)I_3]-2I_2 ,\nonumber \\
f_3^{(1)}\overset{\rm soft}{\sim}&\frac{2(p\cdot r)}{m^2(r\cdot r^*)-2(p\cdot r)(p\cdot r^*)}\bigg\{\bigg[\frac{\omega_p m^2(r\cdot r^*)}{p\cdot r}-3\omega_p(p\cdot r^*)\bigg]I_1+\frac{(p\cdot r^*)^2}{r\cdot r^*}I_2\nonumber \\
&+\bigg[m^2-\frac{(p\cdot r)(p\cdot r^*)}{r\cdot r^*}\bigg]I_3\bigg\} ,\nonumber \\
f_4^{(1)}\overset{\rm soft}{\sim}&\frac{2(p\cdot r)}{m^2(r\cdot r^*)-2(p\cdot r)(p\cdot r^*)}\bigg\{-\omega_p(p\cdot r)I_1+\bigg[m^2-\frac{(p\cdot r)(p\cdot r^*)}{r\cdot r^*}\bigg]I_2\nonumber \\
&+\frac{(p\cdot r)^2}{r\cdot r^*}I_3\bigg\} ,
\end{align}
where
\begin{align}
&I_1=\int \frac{d^Dk}{(2\pi)^D}\frac{1}{k^2[(p-k)^2-m^2](k\cdot r)}\bigg|_{p_0=\omega_p}\overset{\rm soft}{\sim}-\frac{1}{2}\int\frac{d^Dk}{(2\pi)^D}\frac{1}{k^2(k\cdot p)(k\cdot r)}\ ,\nonumber \\
&I_2=\frac{d}{dp_0}\int \frac{d^Dk}{(2\pi)^D}\frac{1}{k^2[(p-k)^2-m^2]}\bigg|_{p_0=\omega_p}\overset{\rm soft}{\sim}-\frac{\omega_p}{2}\int\frac{d^Dk}{(2\pi)^D}\frac{1}{k^2(k\cdot p)^2}\ ,\nonumber \\
&I_3=\frac{d}{dp_0}\int \frac{d^Dk}{(2\pi)^D}\frac{k\cdot r^*}{k^2[(p-k)^2-m^2](k\cdot r)}\bigg|_{p_0=\omega_p}\overset{\rm soft}{\sim}-\frac{\omega_p}{2}\int\frac{d^Dk}{(2\pi)^D}\frac{k\cdot r^*}{k^2(k\cdot p)^2(k\cdot r)}\ .
\end{align}
Note, to regularize the soft divergence by dimensional regularization, the derivative and integral operation should be performed in an order:
\begin{equation}
\frac{d}{dp_0}\int \frac{d^Dk}{(2\pi)^D}\frac{\cdots}{\cdots} \bigg|_{p_0=\omega_p}\to \int \frac{d^Dk}{(2\pi)^D}\left(\frac{d}{dp_0}\frac{\cdots}{\cdots} \bigg|_{p_0=\omega_p}\right).
\end{equation}
Finally, we obtain the desired soft singular term:
\begin{align}
&\frac{1}{2}\bar{u}(p)[-i\Sigma_{c_ic_{i^\prime}}(p)]\frac{i(\slashed{p}+m)}{p^2-m^2}\Gamma_{c_{i^\prime}}(p)\bigg|_{p_0\to \omega_p}\nonumber \\
\overset{\rm soft}{\sim}&\frac{i}{2}g_s^2 C_F\frac{m^2f_1^{(1)}+(p\cdot r)f_3^{(1)}+(p\cdot r^*)f_4^{(1)}}{\omega_p}\bar{u}(p)\Gamma_{c_i}(p)\nonumber \\
\overset{\rm soft}{\sim}&-\frac{i}{2}g_s^2 C_F\mu^{4-D}\int\frac{d^Dk}{(2\pi)^D}\frac{p\cdot \Pi(k)\cdot p}{k^2(k\cdot p)^2}\bar{u}(p)\Gamma_{c_i}(p).
\label{eq_appx_softquark}
\end{align}
The result for massless quark can be obtained by simply set $m=0$, as we have checked that the soft-collinear region will not produce new terms.

The self-energy correction to external gluon is of the form
\begin{equation}
\frac{1}{2}\vcenter{\hbox{\includegraphics[scale=0.5]{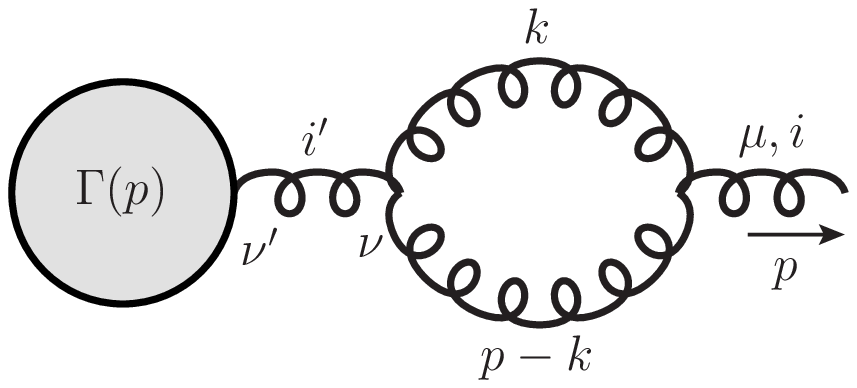}}}=\frac{1}{4}\epsilon^{*\mu}(p)[-i\Sigma^{c_ic_{i^\prime}}_{\mu\nu}(p)]\frac{i\Pi^{\nu\nu^\prime}(p)}{p^2}\Gamma^{c_{i^\prime}}_{\nu^\prime}(p)\bigg|_{p_0\to |\vec{p}|},
\label{eq_appx_gluonsel}
\end{equation}
where
\begin{align}
-i\Sigma_{\mu\nu}^{c_ic_{i^\prime}}(p)=-C_A\delta^{c_ic_{i^\prime}}g_s^2\mu^{4-D}\int\frac{d^Dk}{(2\pi)^D}&\frac{i\Pi^{\alpha\rho}(k)i\Pi^{\beta\sigma}(p-k)}{k^2(p-k)^2}\nonumber \\
&[g_{\rho\sigma}(2k-p)_{\mu}+g_{\sigma\mu}(2p-k)_{\rho}+g_{\mu\rho}(-p-k)_{\sigma}]\nonumber \\
&[g_{\nu\beta}(2p-k)_{\alpha}+g_{\beta\alpha}(2k-p)_{\nu}+g_{\alpha\nu}(-k-p)_{\beta}]\ .
\end{align}
By performing power counting, we obtain the terms that are potentially soft or soft-collinear divergent
\begin{align}
&\frac{1}{4}\epsilon^{*\mu}(p)[-i\Sigma^{c_ic_{i^\prime}}_{\mu\nu}(p)]\frac{i\Pi^{\nu\nu^\prime}(p)}{p^2}\Gamma^{c_{i^\prime}}_{\nu^\prime}(p)\bigg|_{p_0\to |\vec{p}|}\nonumber \\
\overset{\rm soft}{\sim}&-\frac{i}{2}g_s^2C_A\mu^{4-D}\int\frac{d^Dk}{(2\pi)^D}\frac{p\cdot \Pi(k)\cdot p}{k^2(p\cdot k)^2}\Gamma^{c_i}(p)\cdot \epsilon^*(p).
\end{align}
Note, a factor of $2$ is included to incorporate the contribution from $(k-p)\to 0$ region.

\section*{Appendix B: Hard-collinear singularities of external self-energy diagrams}

In this appendix, we present the derivation of Eq. \eqref{eq_collexternal}, the hard-collinear singular terms of external self-energy diagrams.
We decompose the loop momentum as
\begin{equation}
k^\mu=zp^\mu+k^-\bar{p}^\mu+k_T^\mu,
\end{equation}
where $p$ is the external momentum and $\bar{p}=(p_0,-\vec{p})$.
Then the propagator and the integral measures take the forms
\begin{align}
&k^2=4p_0^2zk^--\vec{k}_T^2,\nonumber \\
&(p-k)^2=-4p_0^2(1-z)k^--\vec{k}_T^2,\nonumber \\
&d^Dk=2p_0^2\ dz\ dk^-\ d^{D-2}k_T.
\end{align}
For the quantity $(r\cdot k)^{-1}$, we use the ML prescription as in Appendix A.
To reduce the number of independent vectors, we set $r$ to be parallel with $\bar{p}$ (then $r^*$ is parallel with $p$).
In fact, these treatments is not much concern to the final result, since $(r\cdot k)^{-1}$ is nonvanishing in hard-collinear region (except when choosing $r\varpropto p$).

The self-energy correction to external massless quark is of the form
\begin{equation}
\frac{1}{2}\vcenter{\hbox{\includegraphics[scale=0.5]{softSEQuark.eps}}}=\frac{1}{2}\bar{u}(p)[-i\Sigma_{c_ic_{i^\prime}}(p)]\frac{i\slashed{p}}{p^2}\Gamma_{c_{i^\prime}}(p)\bigg|_{p_0\to |\vec{p}|} ,
\end{equation}
with
\begin{equation}
-i\Sigma_{c_ic_{i^\prime}}(p)=g_s^2C_F\delta_{c_ic_{i^\prime}}\mu^{4-D}\int\frac{d^Dk}{(2\pi)^D}\frac{\gamma^\mu(\slashed{p}-\slashed{k})\gamma^\nu}{k^2(p-k)^2}\Pi_{\mu\nu}(k) .
\end{equation}
In hard-collinear region, the scaling behaviors of lightcone components are $z\sim \kappa^0$, $k^-\sim \kappa^2$, $|\vec{k}_T|\sim \kappa$. By taking the terms which scale like $\mathcal{O}(\kappa^0)$, we have
\begin{align}
\frac{1}{2}\bar{u}(p)[-i\Sigma_{c_ic_{i^\prime}}(p)]\frac{i\slashed{p}}{p^2}\Gamma_{c_{i^\prime}}(p)\bigg|_{p_0\to |\vec{p}|}\overset{\rm coll}{\sim} ig_s^2C_F\mu^{4-D}\int \frac{2p_0^2dzd^{D-2}k_T}{(2\pi)^{D-1}}\frac{(D-10)z+8}{4z}&\nonumber \\
\times\int \frac{dk^-}{2\pi}\frac{1}{[4p_0^2zk^--\vec{k}_T^2+i\varepsilon][-4p_0^2(1-z)k^--\vec{k}_T^2+i\varepsilon]}\bar{u}(p)\Gamma_{c_i}(p) .&
\end{align}
It can be seen that there are two poles located at $k^-=\frac{\vec{k}_T^2-i\varepsilon}{4p_0^2z}$ and $k^-=\frac{\vec{k}_T^2-i\varepsilon}{-4p_0^2(1-z)}$.
In the region $z<0$ or $z>1$, both poles lie in the same half-plane, and no singular term will arise.
In the region $0<z<1$, we have
 \begin{align}
 &\overset{\rm coll}{\sim} -\frac{g_s^2}{4\pi}C_F\mu^{4-D}\int_{\frac{\delta_0}{p_0}}^1dz \frac{(D-10)z+8}{4z}\int_{|\vec{k}_T|<\delta_T}\frac{d^{D-2}k_T}{(2\pi)^{D-2}}\frac{1}{\vec{k}_T^2}\ \bar{u}(p)\Gamma_{c_i}(p)\nonumber \\
 &\overset{\rm coll}{\sim}-\frac{g_s^2}{8\pi^2}C_F\frac{1}{\Gamma(1-\epsilon)}\left(\frac{4\pi\mu^2}{\delta_T^2}\right)^\epsilon \frac{1}{2\epsilon}\left(\frac{3+\epsilon}{2}+2\ln\frac{\delta_0}{p_0}\right)\ \bar{u}(p)\Gamma_{c_i}(p).
 \end{align}
 Here, the cutoff parameter $\delta_0$ is introduced to exclude the soft-collinear region.

 The fermion loop self-energy correction to external gluon takes the form:
 \begin{equation}
\frac{n_{lf}}{2}\vcenter{\hbox{\includegraphics[scale=0.5]{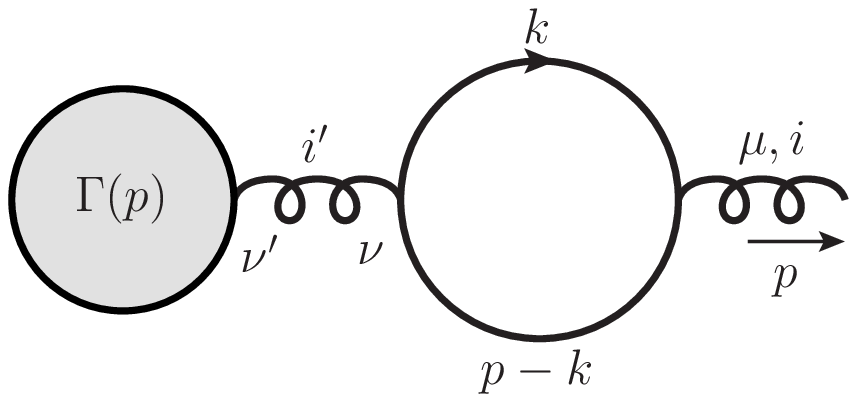}}}=\frac{n_{lf}}{2}\epsilon^{*\mu}(p)[-i\Sigma^{c_ic_{i^\prime}}_{\mu\nu}(p)]\frac{i\Pi^{\nu\nu^\prime}(p)}{p^2}\Gamma^{c_{i^\prime}}_{\nu^\prime}(p)\bigg|_{p_0\to |\vec{p}|}\ ,
\label{eq_appx_gluonsel}
\end{equation}
where
\begin{align}
-i\Sigma_{\mu\nu}^{c_ic_{i^\prime}}(p)=-\frac{g_s^2}{2}\delta^{c_ic_{i^\prime}}\mu^{4-D}\frac{d^Dk}{(2\pi)^D}\frac{{\rm Tr}[\gamma^\nu\cdot (\slashed{k}-\slashed{p})\cdot \gamma^\mu\cdot \slashed{k}]}{k^2(p-k)^2}  .
\end{align}
Similarly, we have
\begin{align}
&\frac{n_{lf}}{2}\epsilon^{*\mu}(p)[-i\Sigma^{c_ic_{i^\prime}}_{\mu\nu}(p)]\frac{i\Pi^{\nu\nu^\prime}(p)}{p^2}\Gamma^{c_i}_{\nu^\prime}(p)\bigg|_{p_0\to |\vec{p}|}\nonumber \\
\overset{\rm coll}{\sim}&i\frac{n_{lf}}{2}g_s^2\frac{D-2}{D-1}\mu^{4-D}\int\frac{d^Dk}{(2\pi)^D}\frac{1}{k^2(p-k)^2}\ \Gamma_b(p)\cdot \epsilon^*(p) \nonumber \\
\overset{\rm coll}{\sim}&-n_{lf}\frac{g_s^2}{8\pi}\frac{D-2}{D-1}\mu^{4-D}\int_{|\vec{k}_T|<\delta_T}\frac{d^{D-2}k_T}{(2\pi)^{D-2}}\frac{1}{\vec{k}_T^2}\ \Gamma_{c_i}(p)\cdot \epsilon^*(p) \nonumber \\
\overset{\rm coll}{\sim}&n_{lf}\frac{g_s^2}{8\pi^2}\frac{1}{\Gamma(1-\epsilon)}\left(\frac{4\pi\mu^2}{\delta_T^2}\right)^\epsilon\frac{1}{2\epsilon}\frac{1-\epsilon}{3-2\epsilon}\ \Gamma_{c_i}(p)\cdot \epsilon^*(p) .
\end{align}

The fermion loop self-energy correction to external gluon takes the form:
\begin{equation}
\frac{1}{2}\vcenter{\hbox{\includegraphics[scale=0.5]{softSEGluon.eps}}}=\frac{1}{4}\epsilon^{*\mu}(p)[-i\Sigma^{c_ic_{i^\prime}}_{\mu\nu}(p)]\frac{i\Pi^{\nu\nu^\prime}(p)}{p^2}\Gamma^{c_{i^\prime}}_{\nu^\prime}(p)\bigg|_{p_0\to |\vec{p}|} ,
\label{eq_appx_gluonsel}
\end{equation}
where
\begin{align}
-i\Sigma_{\mu\nu}^{c_ic_{i^\prime}}(p)=-C_A\delta^{c_ic_{i^\prime}}g_s^2\mu^{4-D}\int\frac{d^Dk}{(2\pi)^D}&\frac{i\Pi^{\alpha\rho}(k)i\Pi^{\beta\sigma}(p-k)}{k^2(p-k)^2}\nonumber \\
&[g_{\rho\sigma}(2k-p)_{\mu}+g_{\sigma\mu}(2p-k)_{\rho}+g_{\mu\rho}(-p-k)_{\sigma}]\nonumber \\
&[g_{\nu\beta}(2p-k)_{\alpha}+g_{\beta\alpha}(2k-p)_{\nu}+g_{\alpha\nu}(-k-p)_{\beta}] .
\end{align}
We have
\begin{align}
&\frac{1}{4}\epsilon^{*\mu}(p)[-i\Sigma^{c_ic_{i^\prime}}_{\mu\nu}(p)]\frac{i\Pi^{\nu\nu^\prime}(p)}{p^2}\Gamma^{c_{i^\prime}}_{\nu^\prime}(p)\bigg|_{p_0\to |\vec{p}|}\nonumber \\
\overset{\rm coll}{\sim}&ig_s^2C_A\mu^{4-D}\int\frac{d^Dk}{(2\pi)^D}\frac{(z^2-z+1)^2}{z(1-z)}\frac{1}{k^2(p-k)^2}\ \Gamma_{c_i}(p)\cdot \epsilon^*(p) \nonumber \\
\overset{\rm coll}{\sim}&-\frac{g_s^2}{4\pi}C_A\mu^{4-D}\int_{\frac{\delta_0}{p_0}}^{1-\frac{\delta_0}{p_0}} dz\frac{(z^2-z+1)^2}{z(1-z)}\int_{|\vec{k}_T|<\delta_T}\frac{d^{D-2}k_T}{(2\pi)^{D-2}}\frac{1}{\vec{k}_T^2}\ \Gamma_{c_i}(p)\cdot \epsilon^*(p) \nonumber \\
\overset{\rm coll}{\sim}&-\frac{g_s^2}{8\pi^2}C_A\frac{1}{\Gamma(1-\epsilon)}\left(\frac{4\pi\mu^2}{\delta_T^2}\right)^\epsilon\frac{1}{2\epsilon}\left(\frac{11}{6}+2\ln\frac{\delta_0}{p_0}\right)\ \Gamma_{c_i}(p)\cdot \epsilon^*(p) .
\end{align}

\end{document}